\documentclass[sigconf]{acmart}
\usepackage{graphicx}
\usepackage{booktabs}
\usepackage{subfigure}

\AtBeginDocument{%
  }

\copyrightyear{2023}
\acmYear{2023}
\setcopyright{acmlicensed}\acmConference[MMAsia '23]{ACM Multimedia Asia 2023}{December 6--8, 2023}{Tainan, Taiwan}
\acmBooktitle{ACM Multimedia Asia 2023 (MMAsia '23), December 6--8, 2023, Tainan, Taiwan}
\acmPrice{15.00}
\acmDOI{10.1145/3595916.3626430}
\acmISBN{979-8-4007-0205-1/23/12}

\acmConference[MMAsia '23]{ACM Multimedia Asia}{December 06--08,
  2023}{Tainan, Taiwan}
\acmPrice{15.00}
\acmISBN{978-1-4503-XXXX-X/18/06}

\begin{document}

\title{RecipeMeta: Metapath-enhanced Recipe Recommendation on Heterogeneous Recipe Network}

\author{Jialiang Shi}
\affiliation{%
  \institution{Nagoya University}
  \city{Aichi}
  \country{Japan}}
\email{shij@cs.is.i.nagoya-u.ac.jp}

\author{Takahiro Komamizu}
\affiliation{%
  \institution{Nagoya University}
  \city{Aichi}
  \country{Japan}}
\email{taka-coma@acm.org}

\author{Keisuke Doman}
\affiliation{%
 \institution{Chukyo University}
 \city{{Aichi}}
 \country{Japan}}
\email{kdoman@sist.chukyo-u.ac.jp}

\author{Haruya Kyutoku}
\affiliation{%
  \institution{Aichi University of Technology}
  \city{Aichi}
  \country{Japan}}
\email{kyutoku-haruya@aut.ac.jp}

\author{Ichiro Ide}
\affiliation{%
  \institution{Nagoya University}
  \city{Aichi}
  \country{Japan}}
\email{ide@i.nagoya-u.ac.jp}

\renewcommand{\shortauthors}{Shi et al.}

\begin{abstract}
Recipe is a set of instructions that describes how to make food. 
It can help people from the preparation of ingredients, food cooking process, etc. to prepare the food, and increasingly in demand on the Web. 
To help users find the vast amount of recipes on the Web, we address the task of recipe recommendation.
Due to multiple data types and relationships in a recipe, we can treat it as a heterogeneous network to describe its information more accurately. 
To effectively utilize the heterogeneous network, metapath was proposed to describe the higher-level semantic information between two entities by defining a compound path from peer entities. 
Therefore, we propose a metapath-enhanced recipe recommendation framework, RecipeMeta, that combines GNN (Graph Neural Network)-based representation learning and specific metapath-based information in a recipe to predict User-Recipe pairs for recommendation. 
Through extensive experiments, we demonstrate that the proposed model, RecipeMeta, outperforms state-of-the-art methods for recipe recommendation.
\end{abstract}




\maketitle
\section{Introduction}
With the growth of the Internet~\cite{tian2022recipe}, people are starting to look for recipes on social media and Websites. For example,  \url{Food.com}\footnote{https://www.food.com/ (Accessed: August 8th, 2023)}, one of the largest
recipe-sharing Websites consists of over half a million recipes. A recipe contains multiple types of information about foods, such as ingredient and instruction.
In addition, recipe sharing services also provide sharing and rating function of recipes. Therefore,
huge amounts of complicated recipe data make it difficult for a user to find desired recipes.
For example, people will pick a satisfied recipe according to their cultures, countries, and ethnicities ~\cite{mokdara2018personalized}. Based on the ingredient and instruction information, there are still plenty of useful information that have not been used effectively~\cite{li2020reciptor}. 

\begin{figure}[t]
  \centering
  \includegraphics[width=0.75\linewidth]{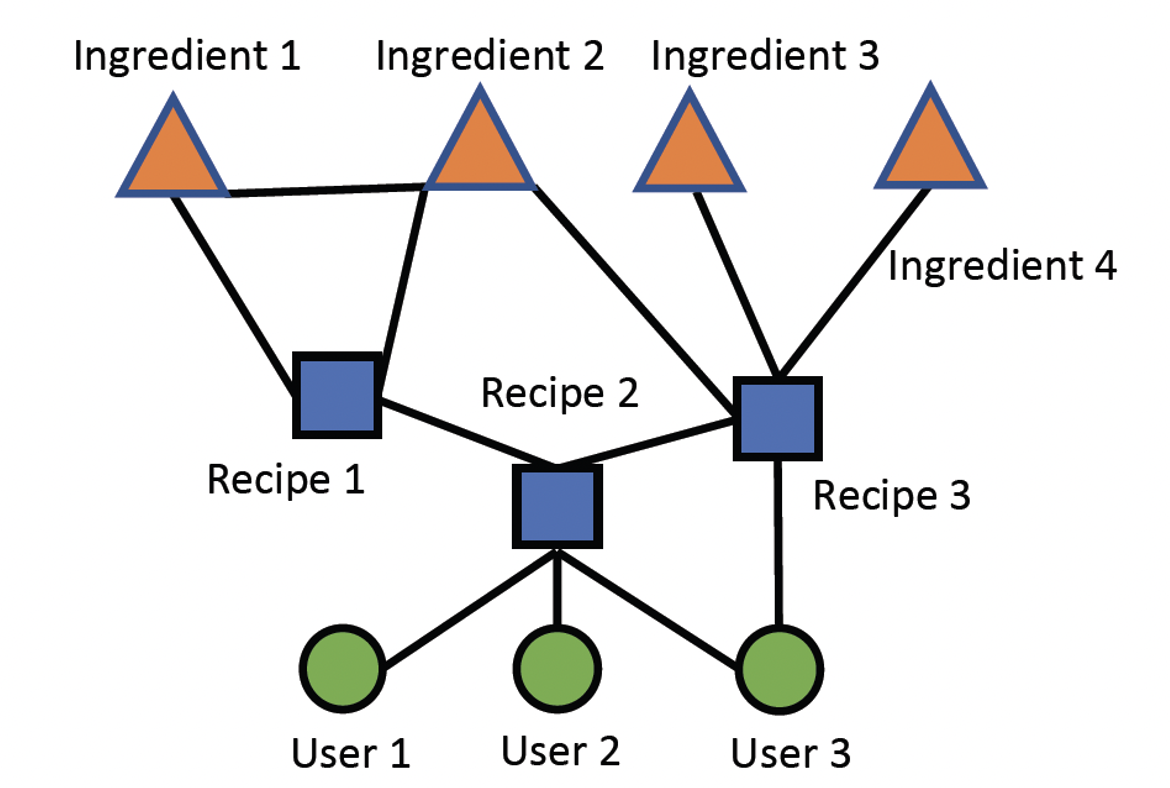}
  \caption{Example of a Heterogeneous Information Network (HeteIN) of recipes. 
  It consists of three types of nodes: User, Recipe, and Ingredient,
  and four types of relations: Recipe-Recipe, User-Recipe, Recipe-Ingredient, and Ingredient-Ingredient.}
  \label{fig:dataset}
\end{figure}

Previous study attempts to use simple aggregation methods to combine the heterogeneous food information like ingredients, instruction, and visual appearance to recommend a recipe ~\cite{gao2019hierarchical,ge2015health}.
However, it cannot accurately take these factors and relationships into account. 
In this paper, we represent recipes and related information as a Heterogeneous Information Network (HeteIN) to capture the relationships among them as shown in Fig.~\ref{fig:dataset}.
In this line, a previous method \cite{tian2022reciperec} constructed and used HeteIN for recipes that treats users, instructions, and ingredients as different node types, while they ignored the differences of their relations. 
However, we consider that these differences are important to predict (hidden) preferences of users.

In a way of fully utilizing HeteIN, \textit{metapath} refers to a specific path in the network schema which is an ordered sequence of node types and edge types~\cite{sun2011pathsim}. 
Previous research~\cite{liu2022survey} suggests that learning high-level semantic information solely by the metapath-based learning mechanism can improve the performance of a recommender system.
However, the limitation is that it requires a large amount of interactive information.
With HeteIN, we can use heterogeneous data to  solve the data sparseness. Then, we combine the heterogeneous data with the rich semantics in metapath for exploring the hidden information. 
\hyphenation{Recipe-Meta}
Based on the idea above, we propose a framework, 
RecipeMeta, to solve the recipe recommendation problem. 
The key idea of RecipeMeta is three-fold: 
(1) We treat the recipe recommendation problem as a link prediction problem,
(2) by following preceding successful researches, Graph Neural Network (GNN) is used for representation learning, and
(3) to enhance semantic relationships among nodes in the recipe HeteIN, a metapath-guided graph method is introduced to reconstruct a heterogeneous network into homogeneous information networks that represent semantically similar relationships through metapath-based similarity calculation.
In particular, first, we define a recipe-related HeteIN consisting of Users, Instructions, Ingredients, and their relationships following \cite{tian2022reciperec}. 
Then, we select several metapaths consisting of specific relations, e.g., for a specific node type Recipe, we select \textbf{Recipe $\rightarrow$ User $\rightarrow$ Recipe} and \textbf{User $\rightarrow$ Recipe $\rightarrow$ User} to compute the similarity for Recipes and Users. 
Based on the similarities between nodes, we reconstruct a homogeneous network between the original nodes and similar nodes,
and use it as the input of GNNs for obtaining metapath-based node information. 
In addition, to fully utilize the HeteIN, another independent GNN-based representation learning is applied to the whole network. 
Finally, these information are combined to perform link prediction.

In summary, our contributions are as follows: 
\begin{itemize}
    \item \textbf{RecipeMeta --- A novel framework}:
        We propose RecipeMeta which is a link prediction-based recipe recommendation framework that combines two representation learning methods on a heterogeneous information network about recipe data; i.e., GNN-based one and metapath-based one.
        It is a convenient and explainable way including metapath to enhance the understanding of each component in a recipe.
    \item \textbf{Effectiveness}: 
        In this paper, the proposed framework, RecipeMeta, is examined by using the URI-Graph~\cite{tian2022reciperec} dataset.
        Experiments show that RecipeMeta outperforms the state-of-the-art recipe recommendation method, RecipeRec~\cite{tian2022reciperec}.
\end{itemize}

\section{Related Work}

Recommendation systems can be categorized into collaborative filtering, content-based recommendation, knowledge-based recommendation, and hybrid recommendation~\cite{jannach2010recommender}. They often face data sparsity and the cold-start problem. Therefore, researchers often use auxiliary information to assist recommendation systems to integrate data globally. Among them, researchers tried to incorporate structured information such as social networks and knowledge graphs. However, these are often specific types of auxiliary information~\cite{yang2014survey}, so it is difficult to understand all the information comprehensively.

In recent years, researchers have tried to use HeteIN to understand the relationship between users and items in the recommendation system. This not only solves the problems of data sparseness and cold start, but also increases the interpretability of the model through the diversity of data. Among other things, researchers found that calculating the similarity of nodes of the same type in a metapath can help the recommendation model effectively explore the potential and structural information~\cite{sun2011pathsim}. Recipes always contain multiple data types and relationships, but the underlying semantic implications included within them have not been emphasized. In this study, we use a basic metapath-based similarity approach by firstly aggregating metapath-based neighbors and then fusing multiple embeddings in a recipe. We combine supervised recommendation information with recipes to achieve effective modeling results than the state-of-the-art~\cite{tian2022reciperec}.


Food recommendation often has their own unique issues: context and knowledge, personal model construction, heterogeneous food analysis, and nutrition and health~\cite{min2019food}. In recent years, due to the rise of social networks and the Web, researchers have tried to incorporate the heterogeneous information present in the recipe into the recommendation model, which resulted in the model not being able to effectively combine the multimodal information to recommend a suitable recipe for a user~\cite{lin2014content}. The emergence of GNN has allowed researchers to connect the different kinds of information in the recipes through graphs~\cite{tian2022reciperec}. Although they have considered HeteIN to merge diverse information with supervised signals, it is difficult to capture the semantic features hidden underneath the recipes in the face of the huge amount of data. In our work, we enhance the learning of the representations by defining various metapaths, which capture the potential high-order information and match users with a more comprehensive understanding of what kind of recipes they want.

\begin{figure*}[t]
  \centering
  \includegraphics[width=0.9\linewidth]{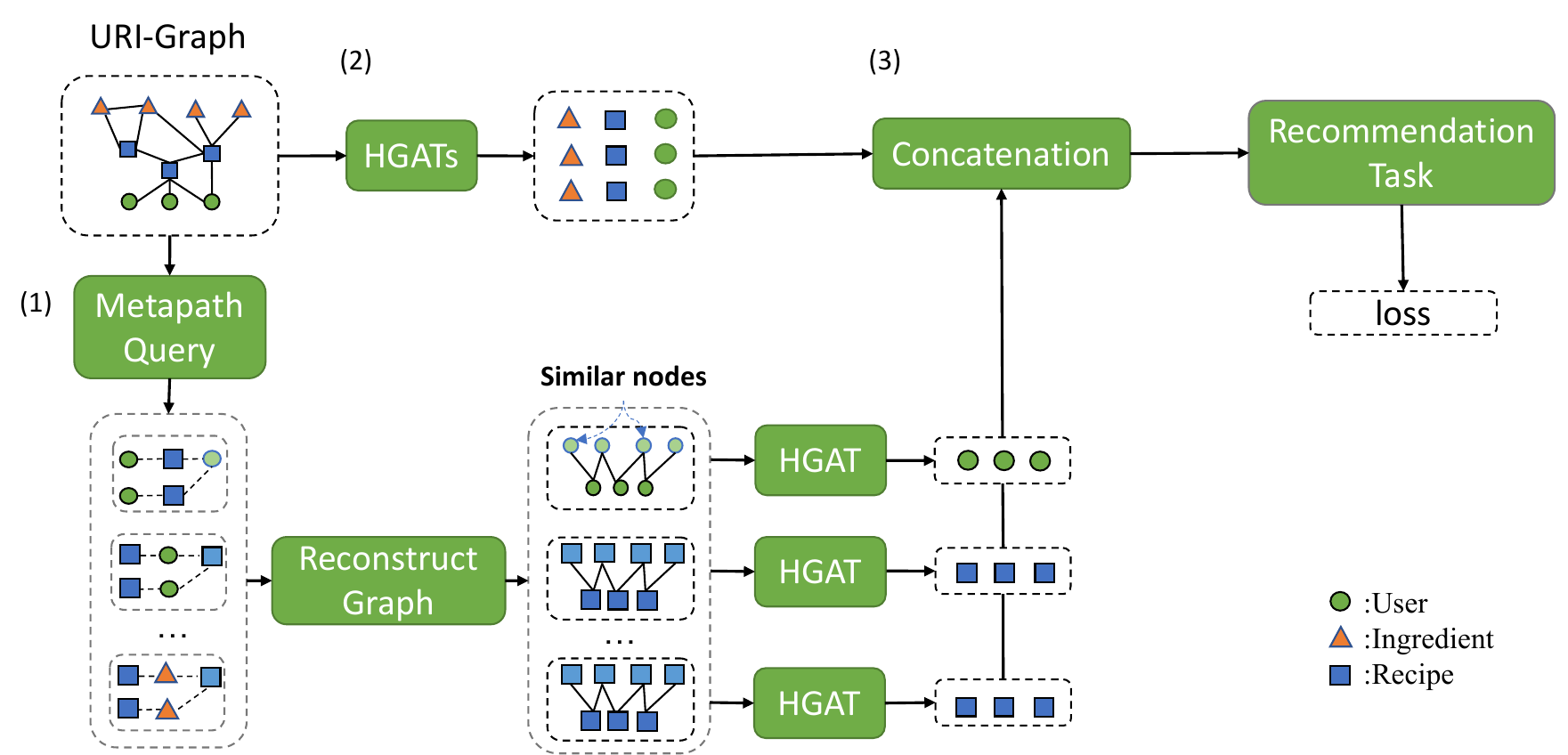}
  \caption{Overall framework with Heterogeneous Graph ATtention networks (HGATs) as the backbone. (1) The first strategy is calculating the similarity according the specific metapath among Recipe and User nodes, and then feed the similar nodes as a graph into HGAT. In the experiment, we take the result of the metapath combination of \textbf{Recipe $\rightarrow$ User $\rightarrow$ Recipe} and \textbf{User $\rightarrow$ Recipe $\rightarrow$ User}  as the baseline result. (2) The second strategy directly feeds the data into another HGAT. (3) Finally, we concatenate the output node embedding and calculate the recipe recommendation loss.}
  \label{fig:framework}
\end{figure*}

\begin{figure}[t]
  \centering
  \includegraphics[width=0.6\linewidth]{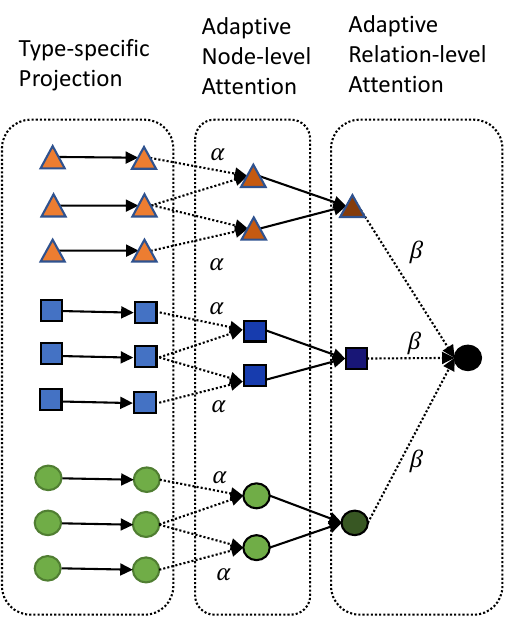}
  \caption{Framework of Heterogeneous Graph ATtention networks (HGATs).}
  \label{fig:HGAT}
\end{figure}

\section{Proposed Framework: RecipeMeta}

\subsection{Task}

In this paper, a HeteIN for recipe-related information is denoted as $G = (V, E, T)$, 
where $T$ represents a set of node types, 
$V = \bigcup_{t \in T} V_t$, such that $\bigcap_{t \in T} V_t = \emptyset$, represents a set of nodes,
and $E \subseteq V \times V$ is a set of relationships between nodes.
In particular, in this paper, $T=\{\mathrm{User}, \mathrm{
Ingredient}, \mathrm{Recipe}\}$, 
the relationship between a Recipe and an Ingredient represents containment,
and that between a User and a Recipe represents preference (more specifically, rating).
Fig.~\ref{fig:dataset} illustrates an example of the HeteIN of recipe data.
Based on the HeteIN, the recipe recommendation task is defined as a link prediction task that 
estimates a missing link between a User and a Recipe based on the present information in the HeteIN.


\subsection{Overview of the Framework}
To address the recommendation task, we develop RecipeMeta, a HeteIN-based link prediction framework with the feature combination of metapath-based similar nodes. 
As illustrated in Fig.~\ref{fig:framework}, RecipeMeta consists of a Heterogeneous Graph ATtention network (HGAT)~\cite{tian2022reciperec} consisting of adaptive node-level attention and relation-level attention module, some single layer HGATs with the graph input constructed by metapath-based similarity of specific nodes and a concatenation layer.

\subsection{HGATs: Heterogeneous Graph ATtention networks}
HGATs~\cite{tian2022reciperec} consists of three parts as shown in Fig.~\ref{fig:HGAT}.
First, the method projects different types of embeddings; User, Recipe, and Ingredient, into the same space: 
\begin{equation}
\mathbf{x}_i = \phi(t) \; \mathrm{s.t.} \; v \in V_t , 
\end{equation}
where $\phi: T \rightarrow \mathbb{R}^d$ is a projection function to a corresponding node $v$ of the given type $t \in T$ in the $d$-dimensional embedding space.

Second, the method integrates neighboring nodes of the same type by the adaptive-node attention~\cite{velivckovic2017graph} from layer $\ell$ to layer $\ell+1$ in GNN. 
We apply a shared weight matrix  $\mathbf{W}_i^\ell \in \mathbb{R}^{d_\ell \times d_{\ell+1}}$ to transform the input features $\mathbf{x}_i$ to $\mathbf{z}_i$ as follows:

\begin{equation}
\mathbf{z}_i = \mathbf{x}_i \cdot \mathbf{W}_{i}^\ell.
\end{equation}

Then, for each layer $\ell$, we calculate the attention score $\alpha^\ell_{ij}$ with a shared weight matrix $\mathbf{W}_{ij}^\ell \in \mathbb{R}^{2d_{\ell+1} \times d_{\ell+1}}$ between $i$ and $j$, where $j$ is the index of a neighbor node $v_j$ of node $v_i$ that directly connects to node $v_i \in V$ through a specific relation. 
We normalize this by the softmax function:
\begin{equation}
\alpha^\ell_{ij} = \mathrm{softmax}\left(\mathbf{W}_{ij}^\ell \cdot (\mathbf{z}_i \parallel \mathbf{z}_j) \right),    
\end{equation}
where $\parallel$ represents the concatenation operation.

We use a weight matrix $\mathbf{W}^\ell \in \mathbb{R}^{d_{\ell} \times d_{\ell+1}}$ and $\alpha_{ij}$ as coefficients, to linearly combine neighboring nodes features with multi-head attention for node$\ i$ in the relation$\ r$:
\begin{equation}
\mathbf{x}^{\ell+1}_{i} = \ \parallel_{m=1}^{M} \mathrm{ReLU}\left(\sum_{j \in N(i)} \alpha^{\ell(m)}_{ij}  \mathbf{z}_j + \mathbf{W}^{\ell(m)}  (\mathbf{x}_i^\ell\odot \mathbf{x}_j^\ell) \right),
\end{equation}
where $M$ is the number of multi-heads,  $N(i)$ is the set of neighbor nodes of $v_i$, and $\odot$ represents the Hadamard product.

Third, the method introduces the adaptive relation-level attention to learn the importance of each relation and fuse all relation-specific node embeddings. In particular, we apply a shared non-linear weight matrix$\ \mathbf{W}_{r} \in \mathbb{R}^{d_{\ell+1} \times d_{\ell+1}}$ to transform the relation-specific node and use the trainable vector $\mathbf{q}$ to calculate the similarities and average it for all node embeddings of a specific relation to obtain the importance score $ w_{i,r}$ for node $i$:
\begin{equation}
w_{i,r} =\frac{1}{|V_r|} \sum_{i \in V_r} \mathrm{tanh}(\mathbf{W}_r \cdot \mathbf{x}^{\ell+1}_{i} + \mathbf{b}) \mathbf{q}^\top,
\end{equation}
where $V_r$ denotes the set of nodes in a specific relation $r$ and $\mathbf{b}$ the bias vector.

We then normalize $w_{i,r}$ to obtain the final relation-level attention weight $\beta_{i,r}$ and and fuse the relation-specific node embeddings $\mathbf{x}^{\ell+1}_{i}$
with it to obtain the final node embedding $\mathbf{x}^L_{i}$:
\begin{eqnarray}
\beta_{i,r} &=& \frac{\exp(w_{i,r})}{\sum_{r \in R} \exp(w_{i,r})}, \\
\mathbf{x}^L_{i} &=& \sum_{r=1}^R \beta_{i,r} \mathbf{x}^{\ell+1}_{i}, 
\end{eqnarray}
where $R$ represents the number of the different relations between User, Recipe, and Ingredient.

\subsection{GNN with Semantics Under Metapath}

To better reflect the semantics under different nodes, we propose a metapath-based GNN model to incorporate the different semantics. Specifically, a metapath is a path defined on the graph of network schema~\cite{sun2011pathsim}: 
Given a schema network $G_S = (T, E_S)$ where $E_S \subseteq T \times T$, 
a metapath$\ P$ is defined as consecutive edges in $G_S$, which can be represented in a more interpretable way as: $t_0 \rightarrow t_1 \rightarrow \dots \rightarrow t_m$, where $t_i \in T$ and $m$ is the length of $P$.
As shown in the bottom part of Fig.~\ref{fig:framework}, first, we compute
the top-$k$ (e.g. $k=10$) similar nodes according to \textit{PathSim}~\cite{sun2011pathsim}, which is a metapath-based similarity measure. 
Given a symmetric metapath $P$ which source and destination vertices are $t \in T$, 
the PathSim-based similarity $o(x, y)$ between two same type instances, $x$ and $y$, of $t$ is as follows:

\begin{equation}
o(x,y) =
    \frac{
        2 \times \left| \{p_{x\sim y}:p_{x\sim y} \in P\} \right|
    }{
        \left| \{p_{x\sim x}:p_{x\sim x} \in P\} \right| +
        \left| \{p_{y\sim y}:p_{y\sim y} \in P\} \right|
    },
\label{eq:pathsim}
\end{equation}

\noindent
where $p_{x\sim y}$ is a path instance of $P$ in $G_S$ between $x$ and $y$,
similarly, $p_{x\sim x}$ (resp. $p_{y\sim y}$) is that between $x$ (resp. $y$).

\begin{figure}[t]
  \centering
  \includegraphics[width=1.075\linewidth]{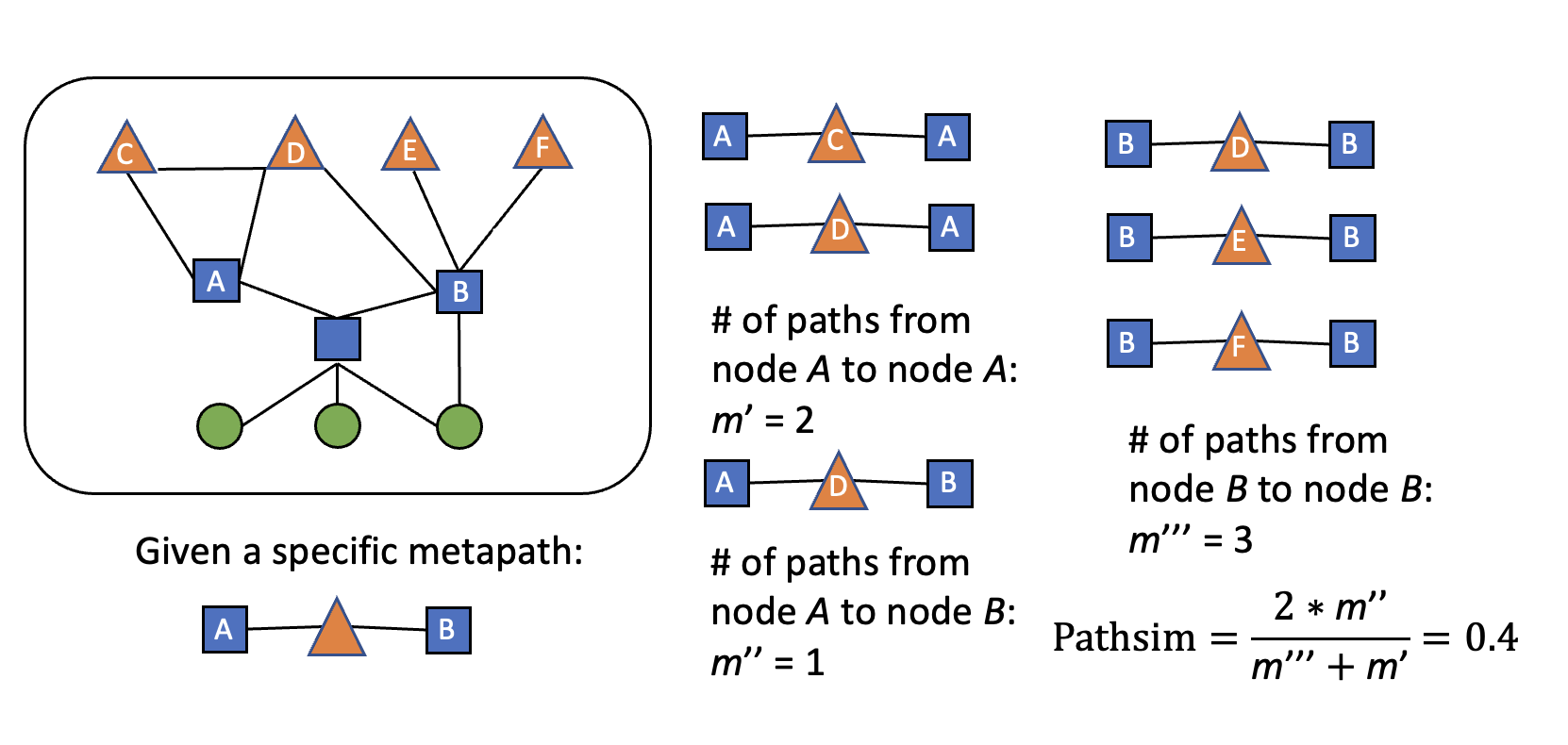}
  \caption{Example of calculating the PathSim score between nodes A and B under a specific metapath.}
  \label{fig:pathsim}
\end{figure}

As an  example in Fig.~\ref{fig:pathsim}, we select a metapath \textbf{Recipe → User → Recipe} and compute $o(A, B)$. There are two paths from node $A$ to node $A$, three paths from node $B$ to node $B$, and one path from node $A$ to node $B$. Finally, we calculate the similarity between nodes $A$ and $B$ as $o(A, B) = \frac{2\cdot m''}{m''' + m'} = \frac{2\cdot 1}{2 + 3} = 0.4$.

Finally, given a symmetric metapath $P$, we form a homogeneous graph $H_P = (V_t, E_P)$ based on the PathSim scores (Eq.~\ref{eq:pathsim}) as follows:
For each node $v \in V_t$ of type $t$ of the starting node of metapath $P$, top-$m$ similar nodes $V'_t$ are extracted, and edges connecting $v$ and all nodes in $V'_t$ are added to $H_P$ such that $E_P \leftarrow E_P \cup \{v\} \times V'_t$.
Then, $H_P$ is fed into a single-layer HGAT to derive the embeddings of nodes of $t$.

\subsection{Objective Function for Recommendation}
In this framework, we want to predict whether a connection exists between a User and a Recipe. Therefore, we adopt the task of link prediction, which can compare the preference scores between nodes connected by a User-Recipe relation in the original graph against the scores between a random pair of User and Recipe node. For example, given a connected User-Recipe pair, we expect the score between User and Recipe would be higher than the score between User and a random Recipe node. Correspondingly, we use an inner product predictor $s(v_u, v_r) = \mathbf{x}_r \cdot \mathbf{x}_r$ to calculate the score between the embeddings of User and Recipe and infer the cross-entropy recommendation loss $L$ as follows: 
\begin{equation}
L = \sum_{v_u \in V_u} \sum_{v_r \in N_r(v_u)} \limits \max \left(0, 1 - s(v_u, v_r) + s(v_u, v'_r)\right) ,
\end{equation}
where $v_u \in V_u$ is a User node, $v_r \in V_r$ are Recipe nodes that is a neighbour of $v_u$ in the User-Recipe relation, and $v'_r \in V_r$ is a randomly selected Recipe node in the HeteIN, respectively.

\section{Experiments for recommendation}

\begin{figure*}[t]
\subfigure[HitRatio@$k$]{
\begin{minipage}[b]{0.31\linewidth}
\includegraphics[width=1\linewidth]{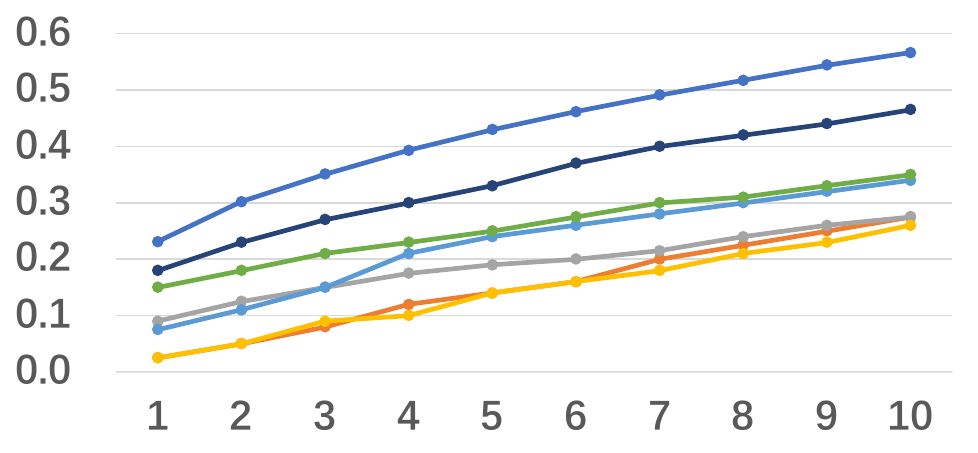}
\end{minipage}}
\subfigure[NDCG@$k$]{
\begin{minipage}[b]{0.31\linewidth}
\includegraphics[width=1\linewidth]{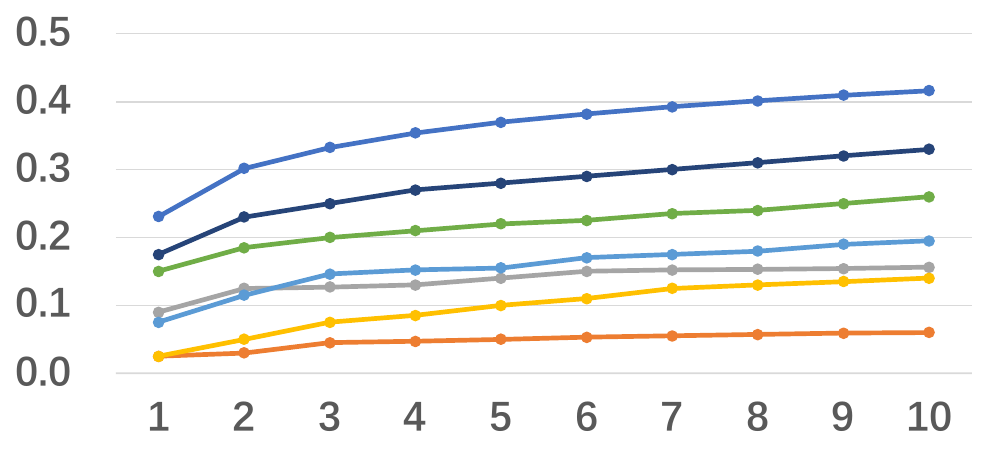}
\end{minipage}}

\subfigure[Precision@$k$]{
\begin{minipage}[b]{0.31\linewidth}
\includegraphics[width=1\linewidth]{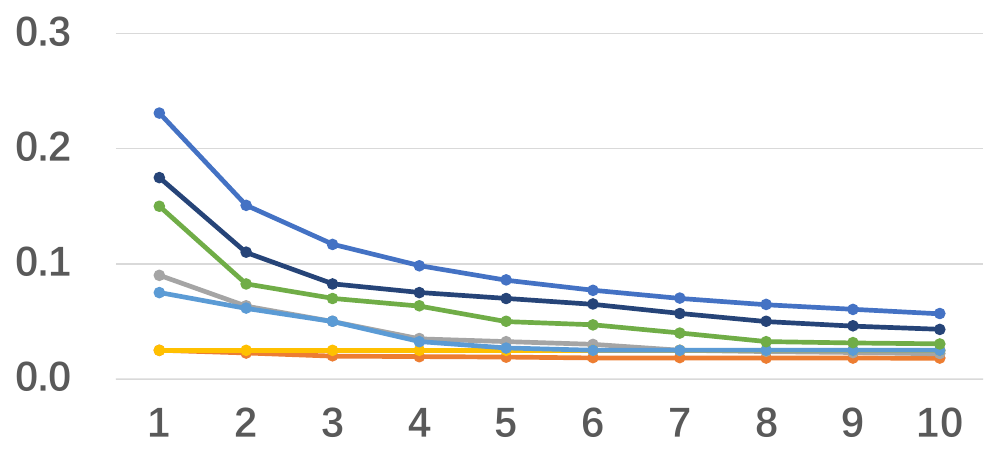}
\end{minipage}}
\subfigure[MAP@$k$]{
\begin{minipage}[b]{0.31\linewidth}
\includegraphics[width=1\linewidth]{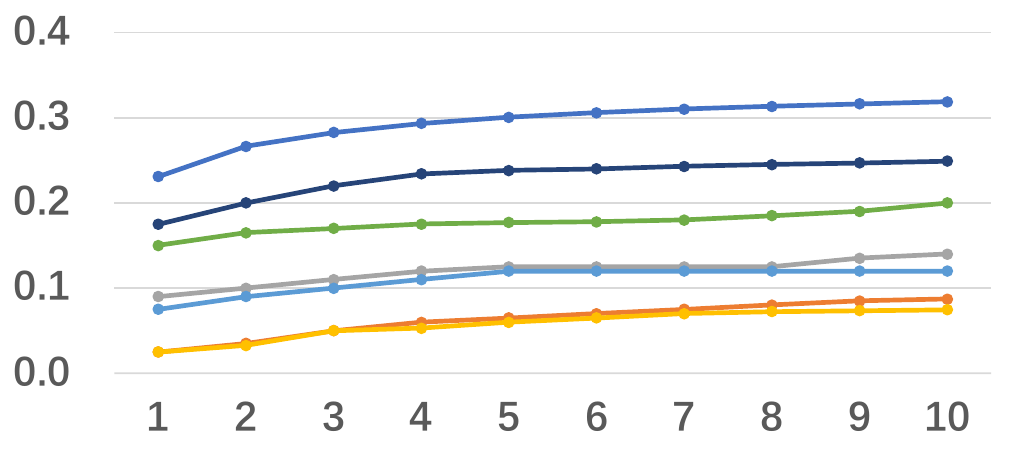}
\end{minipage}}
\includegraphics[width=0.75\linewidth]{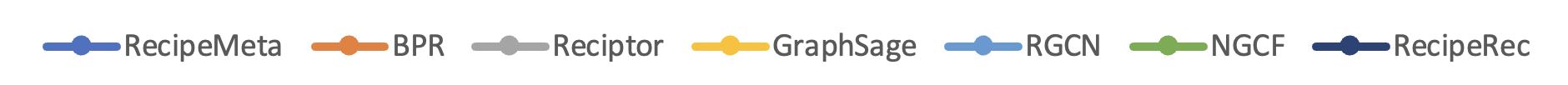}
\caption{Performance of top-$k$ recipe recommendation where $k$ ranges from 1 to 10.}
\label{fig:result}
\end{figure*}

\subsection{Settings}

\paragraph{\bf Dataset}
In this evaluation, URI-Graph~\cite{tian2022reciperec} is used, which is a large-scale User-Recipe-Ingredient heterogeneous network dataset. 
As shown in Fig.~\ref{fig:dataset}, it consists of three different types of nodes, namely, \textbf{User}, \textbf{Ingredient}, and \textbf{Recipe}, and four different types of relations, namely, \textbf{Recipe-Recipe}, \textbf{User-Recipe}, \textbf{Recipe-Ingredient}, and \textbf{Ingredient-Ingredient}. 
The information of Recipe and User was crawled from \url{Food.com}, with 68,794 recipe and 7,958 user nodes in the URI-graph. 
The nutritional information of Ingredient was retrieved from USDA Nutritional Database~\cite{haytowitz2018usda} with 8,847 Ingredient nodes.
Among the relations, Recipe-Recipe relationship represents the degree of similarity between Recipes beyond a threshold where the similarity is calculated by the pre-determined embeddings in the dataset. User-Recipe relationship represents the interactions between User and Recipe with the rating scores. Recipe-Ingredient relationship represents if an Ingredient belongs to a Recipe. Ingredient-Ingredient represents the co-occurring probabilities using Normalized Pointwise Mutual Information~\cite{tian2021recipe}. 

In this experiment, for testing, we randomly split the User-Recipe relationships in the URI-Graph into training / validation / test sets with a ratio of 8:1:1. 
The nodes and edges in the validation and test sets were removed from the URI-graph and the remaining graph was used as the training data.

\paragraph{\bf Evaluation Metric}
To make our results comparable, we followed the same settings as the previous studies~\cite{tian2022reciperec} on URI-graph, whose performance was evaluated by Precision (Pre), Hit Rate (HR), Normalized Discounted Cumulative Gain (NDCG), and Mean Average Precision (MAP) of top-$k$ rankings with $k$ = 1 $\sim$ 10.

\paragraph{\bf Implementation}
For RecipeMeta, we set the learning rate to 0.005, batch size to 412, the training epochs to 50, the maximum number of similar nodes to be retrieved by PathSim to 10. We randomly sampled 100 negative recipes for each user in test phase.

\paragraph{\bf Baseline Methods}
Since our data and task are consistent with the work of RecipeRec~\cite{tian2022reciperec}, we compare the proposed method with the same baselines: 
\begin{itemize}
    \item \textbf{Bayesian Personalized Ranking} $($\textbf{BPR}$)$~\cite{rendle2014bayesian}: A pairwise personalized ranking loss that is derived from the maximum posterior estimator.
    \item \textbf{Reciptor}~\cite{li2020reciptor}: A joint approach for learning effective pretrained recipe
embeddings using both the ingredients and cooking instructions.
    \item \textbf{GraphSAGE}~\cite{hamilton2017inductive}: A framework for inductive representation learning on large graphs.
    \item \textbf{Relational Graph Convolutional Networks} $($\textbf{RGCN}$)$~\cite{schlichtkrull2018modeling}: An application of the GCN framework to modeling relational data.
    \item \textbf{RecipeRec}~\cite{tian2022reciperec}: It can capture recipe content and collaborative signal through a heterogeneous GNN.
    \item \textbf{Neural Graph Collaborative Filtering} $($\textbf{NGCF}$)$~\cite{wang2019neural}: It exploits the user-item graph structure by propagating embeddings on it.
\end{itemize}

\subsection{Performance Comparison}
As Fig.~\ref{fig:result} shows, the proposed RecipeMeta has better results than all baselines over different $k$.
Traditional recommendation algorithm BPR performs the lowest results over the four metrics. For recommendation algorithms designed specifically for recipes such as Reciptor, the results are not optimal because they only learn the representation of Recipe and not other types of nodes. Homogeneous GNNs like GraphSage cannot capture the heterogeneous information either. In comparison to Heterogeneous GNNs like RecipeRec and graph-based recommendation algorithms like NGCF, the proposed RecipeMeta not only considers the integration of different types of information. We also emphasize the semantic paradigm that different types of nodes behave in the graph, such as calculating the PathSim similarity of nodes by choosing metapath. Therefore, the proposed model achieves the best performance, demonstrating its effectiveness.


\subsection{Ablation Study}
\begin{figure*}
\centering
\subfigure[HitRatio@$k$]{
\begin{minipage}[b]{0.23\linewidth}
\includegraphics[width=1\linewidth]{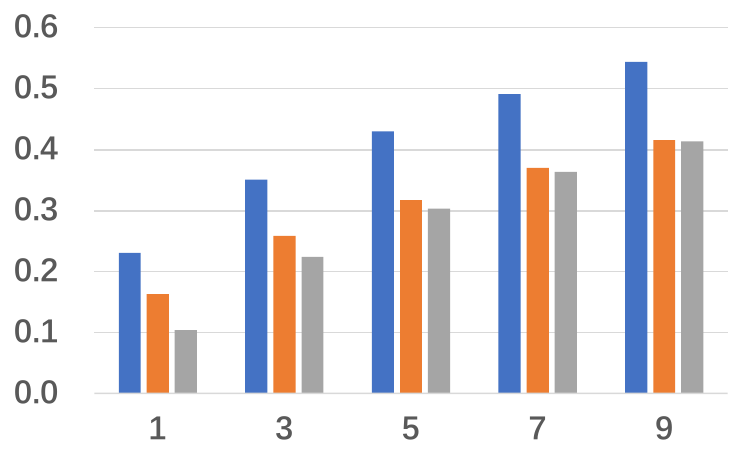}
\end{minipage}}
\subfigure[NDCG@$k$]{
\begin{minipage}[b]{0.23\linewidth}
\includegraphics[width=1\linewidth]{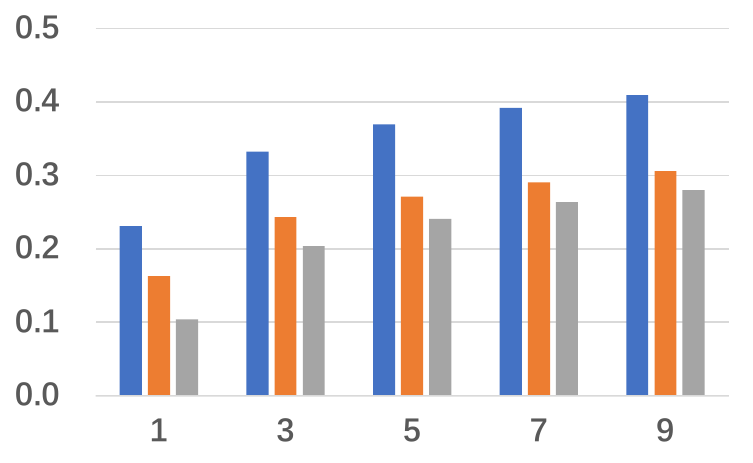}
\end{minipage}}
\subfigure[Precision@$k$]{
\begin{minipage}[b]{0.23\linewidth}
\includegraphics[width=1\linewidth]{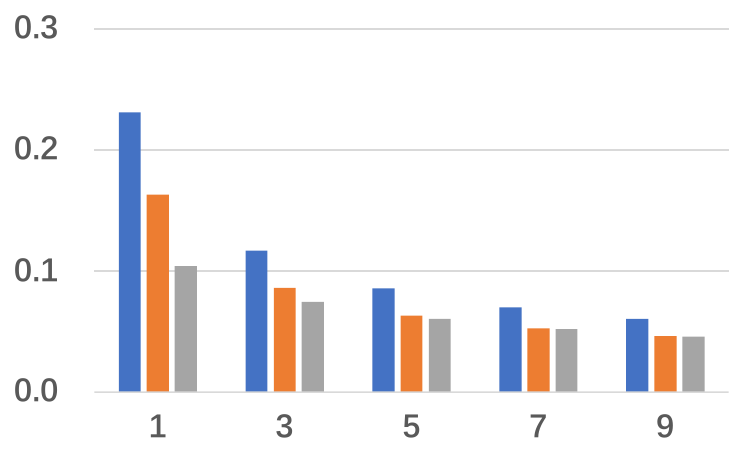}
\end{minipage}}
\subfigure[MAP@$k$
]{
\begin{minipage}[b]{0.23\linewidth}
\includegraphics[width=1\linewidth]{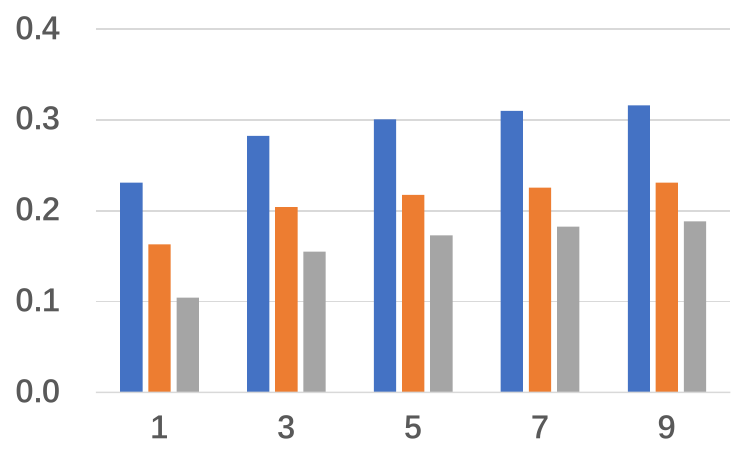}
\end{minipage}}
\includegraphics[width=0.45\linewidth]{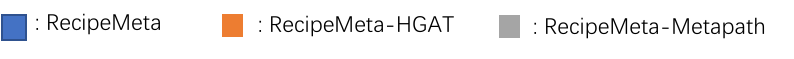}
\caption{Results of different model variant.}
\label{fig:ablation}
\end{figure*}

We conducted ablation studies to evaluate the performance of different model variants including: (a) \textbf{RecipeMeta-HGAT} that only uses HGATs, (b) \textbf{RecipeMeta-Metapath} that only utilizes metapath-based method to reconstruct the input, and (c) \textbf{RecipeMeta} that combines HGATs and metapath-based embeddings. As shown in Fig.~\ref{fig:ablation}, we found that just including HGATs or metapath-based method was no better than a comprehensive model. This also demonstrated that the effectiveness of these modules was realized when the method fused them together.

  

\subsection{Metapath Analysis}
\label{sec:metapath_analysis}
Since the performance of RecipeMeta is partly dependent on the selection of the set of metapaths, we tried to include other metapaths as well. 
To simplify the notation of metapaths, we denote \textbf{U} for user, \textbf{R} for recipe, \textbf{I} for ingredient, and dash (-) for $\rightarrow$.
Since relationships between users were not included in the URI-Graph dataset, we assumed that \textbf{U-R-U} is a fundamental metapath.
In this analysis, we examined the selection of metapaths in two perspectives:
(1) to enhance the sementic relationships between users, and 
(2) to utilize other types of nodes in order to fully utilize the HeteIN.
For (1), we added \textbf{U-R-I-R-U} to enrich the relationships between Users who might be similar if they prefer Recipes including common ingredients.
For (2), we added possible metapaths related to other types (i.e., \textbf{R} and \textbf{I}) than Users, namely, \textbf{R-U-R}, \textbf{R-I-R}, and \textbf{I-R-I}.

Table~\ref{tab:metapath} shows the RecipeMeta results of different sets of metapaths, which indicates that adding different kinds of metapaths, the model can capture different effects. Although adding the longer
metapath \textbf{U-R-I-R-U} showed a positive effect, it was limited in comparison with the other metapaths (like \textbf{R-U-R} and \textbf{R-I-R}). The combination, \{\textbf{U-R-U}, \textbf{R-U-R}, \textbf{R-I-R}\}, was the most effective. 
Meanwhile, adding \textbf{I-R-I} degraded the performance.
These facts indicate that a longer metapath may lose semantics,
and we should focus on the most relevant types of nodes for recipe recommendation, i.e., Recipe and User.

\begin{table}[t]
\centering
\caption{
Average effect of different sets of metapaths on recommendation metrics.
U, R, and I refer to User, Recipe, and Ingredient, respectively.
}
\resizebox{0.45\textwidth}{!}{%
\begin{tabular}{lcccc}
\toprule
Set of Metapaths                & HitRatio       & NDCG      & Precision            & MAP           \\
\midrule
\{U-R-U\}             & 0.409 &        0.341   &   0.096       &  0.276   \\
+ \{U-R-I-R-U\} &     0.420    &0.349  &0.097  & 0.278  \\
\midrule
\{U-R-U\}             & 0.409 &        0.341   &   0.096       &  0.276   \\
+ \{R-U-R\}             & 0.429 & 0.359          & 0.101          & 0.294          \\
+ \{R-U-R, R-I-R\} & \textbf{0.438}          & \textbf{0.367} & \textbf{0.104} & \textbf{0.303} \\
+ \{R-U-R, R-I-R, I-R-I\} &  0.422     & 0.352 & 0.100 & 0.290 \\

\bottomrule
\end{tabular}%
}
\label{tab:metapath}
\end{table}

\subsection{Parameter Sensitivity}
In this experiment, we attempted to reveal the effect of the number $m$ of similar nodes per source node when constructing the homogeneous networks. 
Based on the Metapath Analysis in Section~\ref{sec:metapath_analysis}, 
we chose the combination of metapaths \textbf{U-R-U} and \textbf{R-U-R}.
Table~\ref{tab:parameter} shows the performances of RecipeMeta for different $m$.
Generally speaking, we found that as $m$ increased, the recommendation performance improved. Since the number of similar nodes represents the similar ability between one and other nodes, with a higher $m$, the target node will be capable of finding its own semantics through more similar nodes.
Although all metrics are highest when $m$ = 8, this experiment was limited in a certain range of $m$.
For future experiments, we will examine much higher $m$ values to observe how dense the homogeneous networks should be for better metapath-based GNN representation learning.

\begin{table}[t]
\centering
\caption{Average performance variation of the number $m$ of similar nodes according to the metapath calculation.}
\begin{tabular}{r|cccc}
\hline
$m$ & HitRatio        & NDCG         & Precision               & MAP              \\
\hline
1             & 0.3326      & 0.2712        & 0.0764       & 0.2173      \\ 
2             & 0.3553        & 0.2892      & 0.0818       & 0.2334       \\
3             & 0.3858        & 0.3198       & 0.0904   & 0.2613       \\ 
4             &0.4113        & 0.3438     & 0.0971      & 0.2823         \\ 
5             &  0.4139                 &           0.3412        &      0.0961             &      0.2747             \\ 
6             &     0.4117              &    0.3387              &  0.0950                 &    0.2698               \\ 
7             &0.4369                  &0.3617                   &0.1019                   &0.2926                   \\ 
8             &\textbf{0.4542}                   &\textbf{0.3796}                   &\textbf{0.1068}                   &\textbf{0.3087}                   \\ 
9             &0.4420                   & 0.3687                   & 0.1037                  & 0.2990                  \\ 
10            & 0.4286 & 0.3590 & 0.1012 & 0.2938 \\ 
\hline
\end{tabular}%
\label{tab:parameter}
\end{table}


\section{Conclusion}

In this paper, we proposed a framework, RecipeMeta which combined the metapath-based learning method with GNN to successfully learn the potential relation information of different node types in a recipe for recommendation. Particularly, the method selected specific metapaths and reconstructed a graph with specific similar nodes, and input it to GNN to predict User-Recipe pairs by link prediction. There are still many other metapath-based similarity computation methods, and the emergence of the novel data type on the social media and Web also provides opportunities for future research. We believe that in the future, by integrating the multiple heterogeneous information effectively, we will be able to capture the diversity of data and recommend more satisfying recipes for users.

\subsection*{Acknowledgments}
Parts of the work presented in this paper were supported by the JSPS Grants-in-aid for Scientific Research (21H03555 and 22H00548).

\bibliographystyle{ACM-Reference-Format}
\bibliography{main.bib}

\end{document}